\documentstyle[aps,prl,twocolumn,epsfig]{revtex}  

\font\schwell=schwell at 9pt
\renewcommand{\l}{\schwell L\/}

\title{\large \bf Charmonium dissociation in hadronic matter}
\author{Kevin L.~Haglin\thanks{Email: haglin@stcloudstate.edu}}
\address{Department of Physics, Astronomy and Engineering Science\\
         Saint Cloud State University\\
         St. Cloud, MN 56301, USA}
\date{\today}

\begin{document}


\maketitle

\begin{abstract}
Production of {\it J\/}/$\psi$ from nucleus-nucleus reactions depends 
sensitively on the dissociation cross section with light hadrons.  
Effective lagrangian methods are used to describe the hadronic degrees of
freedom, including strangeness and charm.  Cross sections with pions, rho 
mesons, kaons and nucleons having magnitudes 4--8 mb are found, and with steep
thresholds.  This, folded with thermal momentum distributions for the 
scattering partners, suggests a mean dissociation lifetime 
$\simeq$ 20 fm/{\it c\/}.  Therefore, the ``abnormal'' {\it J\/}/$\psi$
suppression seen in recent Pb+Pb experiments seems to owe to expected 
hadron kinetics.
\end{abstract}

\pacs{PACS numbers: 25.75.-q, 13.75.Lb, 14.40.Lb}

Response of nuclear matter to high energy densities affords the
possibility of creating in the laboratory a system of
quantum chromodynamic (QCD) matter,
the so-called quark gluon plasma.   Among the signatures for 
creation of the mesoscopic colored volume is an idea put forward in 1986 
by Matsui and Satz\cite{MaSa86} to look at electromagnetic spectra for 
evidence of charmonium, bound states of charm-anticharm.  They are very tightly
bound and consequently relatively small hadrons which ought to effectively
probe superdense hadronic matter.  Their utility in this 
context comes from the fact that Debye [color] screening in the plasma 
would so strongly suppress 
{\it c\/}$\bar{\it c\/}$ binding in favor of charm propagating 
decoupled from anticharm to later join 
with more abundant light anti-quark species forming {\it D\/} mesons, that 
suppression of {\it J\/}/$\psi$ would indicate plasma formation.
Above some critical temperature,
the screening radius becomes smaller than the binding radius offering
this possibility.  {\it J\/}/$\psi$ suppression has since been regarded
as a promising signature of quark gluon plasma formation.

Charmonium production cross sections from proton-induced reactions that 
are directly proportional to the target mass number {\it A\/}, are said to 
be normal.  And yet, 
proton-nucleus and nucleus-nucleus experiments performed over the
past several years have exhibited
a common {\it A\/}$^{\alpha}$ dependence, with $\alpha$ $\approx$ 
0.92\cite{E772_91,NA38_91,NA38_95}.   This suppression is now understood
as being due to absorption of the precursor state to {\it J\/}/$\psi$ on 
nucleons and so is in some sense of normal hadronic consequences.
Additional suppression of $\psi^{\prime}$ production revealed in 
nucleus-nucleus experiments has been attributed to
absorption on produced comoving hadrons\cite{GaVo,Wo96}.
However, recent experimental studies at CERN have investigated 
{\it J\/}/$\psi$ production in Pb+Pb reactions at 158 
GeV/nucleon and have reported a dramatic ``abnormal'' suppression compared 
to lighter systems' ({\it AB\/})$^{\alpha}$ behavior\cite{NA50_96}.  It 
is reasonable to
begin discussing the possibility of deconfinement being responsible.
But first, all conventional suppression mechanisms must be under control.

Absorption of {\it J\/}/$\psi$ or $\psi^{\prime}$ on comovers has
been proposed as the dominant dissociating effect.  One mechanism owes to 
deconfinement; if the plasma were present, charmonium states would not form 
at all due to screening.  This is simply a restatement of the signature idea 
of Matsui and Satz.  Another mechanism discussed in the literature is 
pre-thermal dissociation, where the charmonium is excited above the 
({\it D\/}$\bar{\it D\/}$) threshold either by a partonic medium 
which has not had sufficient time to equilibrate\cite{Xu_96}, or by color 
flux tubes in the infant QCD medium\cite{Ne_89,Lo_97}.  The effectiveness of 
pre-thermal suppression mechanisms rests on separation of time scales: the 
first mechanism requires that dissociation proceeds faster than thermalization 
and the second requires that color flux tubes decay more slowly than 
dissociation.
Finally, we come to thermal dissociation where the ({\it 
c\/}$\overline{\it c\/}$) is absorbed through one of the myriad processes 
involving a light hadron.  

The crucial question is the magnitudes of 
cross sections for {\it J\/}/$\psi$ + {\it h\/},
where {\it h\/} is one of
the set ($\pi$, $\rho$, {\it K\/}, {\it N\/}\ldots).   Estimates up
to now have included 1) a perturbative approach at the quark level
where light hadrons interact with {\it J\/}/$\psi$ solely through 
gluonic content effects\cite{Pe_79,Bh_79}, 2) a nonperturbative study
with quark exchange including a confining potential\cite{Mar_95}, and
finally 3) an effective lagrangian approach\cite{Ma_98}.
When the perturbative approach of Peskin and Bhanot was applied by
Kharzeev and Satz to pions 
interacting with {\it J\/}/$\psi$, a cross section $\stackrel{<}{\sim}$
0.1 mb was found for $\sqrt{s}$ roughly 1 GeV above threshold\cite{Kh_94}.
Results for the nonperturbative calculation of Martins, Blaschke and
Quack applied to pions revealed cross sections peaking at several mb also
about 1 GeV above threshold.  The hadronic field theory calculation
of Matinyan and M\"uller resulted in cross sections with pions and
rho mesons of order 1 mb for similar energies.  This situation is
unsettling as there is roughly two orders of magnitude discrepancy
in these estimates.   The aim of this letter is to report on further
investigation of the
issue using effective lagrangian methods quantifying cross sections
within a consistent gauge invariant treatment (including
contact terms which are missing in previous analyses) and including a 
more complete hadronic medium as input to kinetic theory for a baseline 
estimate of the dissociation rate in hadronic matter. 

Hot hadronic matter is populated most abundantly by $\pi$, {\it K\/},
and $\rho$\cite{HaPr_94}.  If circumstances of an appreciable baryon 
chemical potential arise, the nucleons become important as well.  Each 
species can induce charmonium dissociation with respective final states 
governed by conservation laws.  A consistent treatment of light mesons, 
heavy mesons and possibly
baryons is needed to work toward a reasonably reliable description for 
hadronic matter.  We begin with an {\it SU\/}(4) symmetry
so as to include charm, and we
introduce pseudoscalar ({\it P\/} = $\varphi_{a}\lambda_{a}$) and
vector ({\it V\/}$^{\mu}$ = ${v\/}_{a}^{\mu}\lambda_{a}$) meson
matrices, where $\varphi_{a}$ and ${v\/}_{a}^{\mu}$ are pseudoscalar
and vector multiplets and the $\lambda$s are {\it SU\/}(4) 
generators\cite{Kaku}.  The symmetry is severely broken due to the
large charm quark mass, so we use physical mass eigenstates and
matrices, and incorporate constraints where possible.

The free meson lagrangian reads
\begin{eqnarray}
\mbox{\l}_{\,0} & = & \mbox{\rm Tr}(\partial^{\mu}{\it 
P\/}^{\,\dag}\partial_{\mu}
{\it P\/})\nonumber\\
& \ &  - \frac{1}{2}
\mbox{\rm Tr}(\partial^{\mu}{\it V\/}^{\,\dag,\,\nu}-
\partial^{\nu}{\it V\/}^{\,\dag,\,\mu})(\partial_{\mu}{\it V\/}_{\nu}
-\partial_{\nu}{\it V\/}_{\mu}) \nonumber\\
& \ & + \mbox{\rm mass\ terms}.
\end{eqnarray}
Interactions are generated by replacing the spacetime derivative 
with a gauge covariant one $\partial_{\mu}\mbox{\it f\/}\rightarrow 
\mbox{\it D\/}_{\mu}\mbox{\it f\/\rm =} \partial_{\mu}\mbox{\it f\/\rm + 
[{\it A\/}}_{\mu}\mbox{\rm ,\it f\,\rm{]}}$, 
where  $\mbox{\it A\/}_{\mu}$ = -$ig/2\,V_{\mu}$.  The 
vector mesons are recognizable
as playing roles of gauge bosons.  As is usual in effective field theory
strategies, to keep gauge consistency, we must collect terms of 
order $g^{\,2}$.  They are (using {\it P\/}$^{\dag}$ = {\it P\/}
and {\it V\/}$^{\dag}$ = {\it V\/})
\begin{eqnarray}
\mbox{\l}_{\rm\,int} & = & ig\,\mbox{\rm Tr}({\it P\/}{\it 
V\/}^{\mu}
\partial_{\mu}{\it P\/}-\partial^{\mu}{\it P\/}{\it V\/}_{\mu}{\it
P\/}) \nonumber\\
& \ & +\frac{1}{2}\,g^{\,2}\,\mbox{\rm Tr}({\it P\/}{\it V\/}^{\mu}{\it 
V\/}_{\mu}{\it P\/}-{\it P\/}{\it V\/}^{\mu}{\it 
P\/}{\it V\/}_{\mu}) \nonumber\\
& \ & +\,ig\,\mbox{\rm Tr}\left(\partial^{\mu}{\it V\/}^{\nu\,}
\lbrack
{\it V\/}_{\mu}, {\it V\/}_{\nu}
\rbrack 
+ 
\lbrack
{\it V\/}^{\mu\,}, {\it V\/}^{\nu\,}
\rbrack 
\partial_{\mu}{\it V\/}_{\nu}\right) \nonumber\\
& \ & +\,{g^{\,2}}\mbox{\rm Tr}\left(
{\it V\/}^{\mu\,}
{\it V\/}^{\nu\,}
\left\lbrack{\it V\/}_{\mu}, {\it V\/}_{\nu}\right\rbrack\right).
\end{eqnarray}
The order $g$ terms deliver three-point vertices and the order
$g^{\,2}$ terms are responsible for so-called contact terms, or
four-point couplings which are necessary for a gauge invariant
theory.  The interactions of interest for this study are the following
(using $\psi$ as shorthand for $J/\psi$)
\begin{eqnarray}
\label{lpdds}
\mbox{\l}_{\,\pi{D}{D}^*} & = & \frac{i}{2}\,g_{\pi\,D\,D^*}\left(
\bar{\it D\/}\tau_{i}{\it D\/}^{*\,\mu}\partial_{\mu}\pi_{i} 
- \partial^{\mu}\bar{\it D\/}\tau_{i}{\it D\/}_{\mu}^{*}\pi_{i}\right.
\nonumber\\
& \ &  - \left.
\mbox{\rm H.c.}\right), \nonumber\\
\mbox{\l}_{\,\rho{D}{D}} & = & \frac{i}{2}\,g_{\rho{D}{D}}\rho^{\mu}_{i}\left(
\bar{\it D\/}\tau_{i}\partial_{\mu}{\it D\/}
- \partial_{\mu}\bar{\it D\/}\tau_{i} {\it D\/}\right), \nonumber\\
\mbox{\l}_{\,\rho\,D^{*}D^{*}} & = & 
-\frac{i}{2}\,g_{{\rho}D^{*}D^{*}}\rho^{\mu}_{i}\left(
\bar{\it D\/}^{*\,\nu}\tau_{i}(\partial_{\mu}{\it D\/}_{\nu}^{*})
- (\partial_{\mu}\bar{\it D\/}^{*\,\nu})\tau_{i}{\it D\/}_{\nu}^{*} \right.
\nonumber\\
& \ & \left.
- (\partial_{\nu}{\it D\/}^{*}_{\mu})\tau_{i}\bar{\it D\/}^{*\,\nu}
+ (\partial_{\nu}\bar{\it D\/}^{*}_{\mu})\tau_{i}{\it D\/}^{*\,\nu}
\right), 
\nonumber\\
\mbox{\l}_{\,\psi{D}{D}} & = & ig_{\psi{D}{D}}\psi^{\mu}\left(
\bar{\it D\/}\partial_{\mu}{\it D\/}
- (\partial_{\mu}\bar{\it D\/}){\it D\/}\right), \nonumber\\
\mbox{\l\/}_{\,\psi{D}^{*}{D}^{*}} & = & 
-ig_{\psi\,D^{*}D^{*}}\psi^{\mu}\left(
\bar{\it D\/}^{*\,\nu}(\partial_{\mu}{\it D\/}_{\nu}^{*})
- (\partial_{\mu}\bar{\it D\/}^{*\,\nu}){\it D\/}_{\nu}^{*} \right.
\nonumber\\
& \ & \left.
- (\partial_{\nu}{\it D\/}^{*}_{\mu})\bar{\it D\/}^{*\,\nu}
+ (\partial_{\nu}\bar{\it D\/}^{*}_{\mu}){\it D\/}^{*\,\nu}
\right), 
\end{eqnarray}
and
\begin{eqnarray}
\mbox{\l}_{\,\psi{\,\pi}{D}{D}^{*}} & = & 
-g_{\psi{D}{D}}
\,g_{\pi{D}D^{*}}
\psi^{\mu}{\it D\/}^{*}{_\mu}\tau_{i}
\bar{\it D\/}\pi_{i}, \nonumber\\
\mbox{\l}_{\,\psi{\rho}{D}{D}} & = & 
-g_{\rho{D}{D}}
\,g_{\psi{D}{D}}\psi^{\mu}\rho_{\mu\,,i}
\bar{\it D\/}\tau_{i}{\it D\/}, \nonumber\\
\label{lprdsds}
\mbox{\l}_{\,\psi{\rho}{\it D\/}^{*}{\it D\/}^{*}} & = & 
\frac{1}{2}\,g_{\psi{\it D\/}^{*}{\it D\/}^{*}}
\,g_{\rho{\it D\/}^{*}{\it D\/}^{*}}\psi^{\mu}
\left(2\rho_{\mu\,,i}\bar{\it 
D\/}^{*\,\nu}\tau_{i}{\it D\/}^{*}_{\nu}
\right.
\nonumber\\ 
& \ & 
\left.
-\rho^{\nu}_{i}\bar{\it D\/}^{*}_{\mu}\tau_{i}{\it D\/}^{*}_{\nu} 
-\rho^{\nu}_{i}\bar{\it D\/}^{*}_{\nu}\tau_{i}{\it D\/}^{*}_{\mu} 
\right).
\end{eqnarray}
The {\it K\/}-{\it D\/}$_{s}$-{\it D\/}$^{*}$ interaction has the same
structure as the 
$\pi$-{\it D\/}-{\it D\/}$^{*}$, complete with contact terms.

\begin{figure}[htbp]
  \begin{center}
  \epsfxsize 80mm 
  \epsfbox{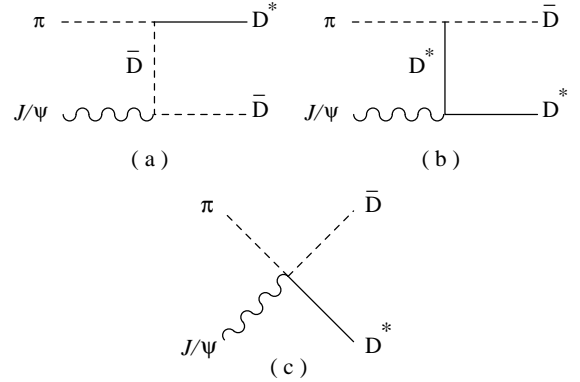}
    \caption{Feynman graphs for the process {\it J\/}/$\psi$ + $\pi
             \rightarrow$ $\bar{\it D\/}$ + {\it D\/}$^{*}$.
            }
    \label{f1}
  \end{center}
\end{figure}

In this approach there are several coupling constants in
Eqs.~(\ref{lpdds}) and (\ref{lprdsds}).
Methods for constraining them will be 
discussed below.  The ${\cal O\/}$({\it g$^{\,2}$}) 
terms carry one power of coupling constant for each 
three-point vertices from which the
contact term collapses. 
As a specific example, the direct, exchange, and 
contact graphs for reaction {\it J\/}/$\psi$ + $\pi 
\rightarrow$ {\it D\/}$^{*}$ + 
$\bar{\it D\/}$ are shown in Fig.~\ref{f1} (a), (b) and (c), respectively.
Contribution to the amplitude from each graph is proportional  
to ${\it g\/}_{\pi\,D\,D^{*}}{\it g\/}_{\psi\,D\,D}$.

Appealing to measured hadronic decays to constrain the coupling
constants gains very little in this case as the best measurement of
the {\it D\/}$^{*\,\pm}$ width has an upper limit of 131 keV\cite{accmor}. 
Model calculations based on relativistic potential
approaches suggest $\Gamma_{{\it D\/}^{*\,\pm}}$ = 46 keV\cite{Co94}.  This 
corresponds to a value for {\it g\/}$_{\pi{\it D\/}{\it D\/}^{*}}$ of 8.8, 
which will be used here.  In the absence of empirical constraints, all other 
coupling constants are obtained from vector meson dominance or flavor 
symmetry arguments.  Vector dominance gives\cite{Ma_98}
\begin{eqnarray}
g_{\rho\,DD} = g_{\rho\,D\/^{*}D\/^{*}} = 5.6, 
\quad
{\it g\/}_{\psi{D}{D}} = {\it g\/}_{\psi{D\/}^{*}{D\/}^{*}} = 7.7.
\end{eqnarray}
For the kaons, we use {\it SU\/}(4) symmetry to relate 
{\it g\/}$_{K{\it D\/}_{s}{\it D\/}^{*}}$ =
$\sqrt{2}\,${\it g\/}$_{\rho\pi\pi}$ =  8.5, and the $\rho\pi\pi$ 
coupling is adjusted to give a $\rho$ width of 151 MeV.

The following absorption reactions are considered
\begin{eqnarray}
\label{piplusjpsi}
\pi + {\it J\/}/\psi &\rightarrow & {\it D\/} + \bar{\it D\/}^{*},
\quad \bar{\it D\/} + {\it D\/}^{*};
\\
\label{rhoplusjpsi}
\rho + {\it J\/}/\psi &\rightarrow & {\it D\/} + \bar{\it D\/},
\quad \bar{\it D\/}^{*} + {\it D\/}^{*};
\\
{\it K\/} + {\it J\/}/\psi &\rightarrow & {\it D\/}_{s} + \bar{\it D\/},
\quad \bar{\it D\/}_{s} + {\it D\/},\nonumber\\
\label{kaplusjpsi}
& \ &
\quad \bar{\it D\/} + {\it D\/}_{s}^{*},
\quad {\it D\/} + \bar{\it D\/}_{s}^{*}.
\end{eqnarray}
Just as in Fig.~\ref{f1}, for a reaction of type
(\ref{piplusjpsi}), each of the reactions listed in
(\ref{piplusjpsi})--(\ref{kaplusjpsi})
has a direct, exchange and a ``seagull'' graph contributing to the amplitude.
Full expressions for the amplitudes and other details will be published 
elsewhere\cite{kh99}.  

\begin{figure}[htbp]
  \begin{center}
  \epsfxsize 80mm 
  \epsfbox{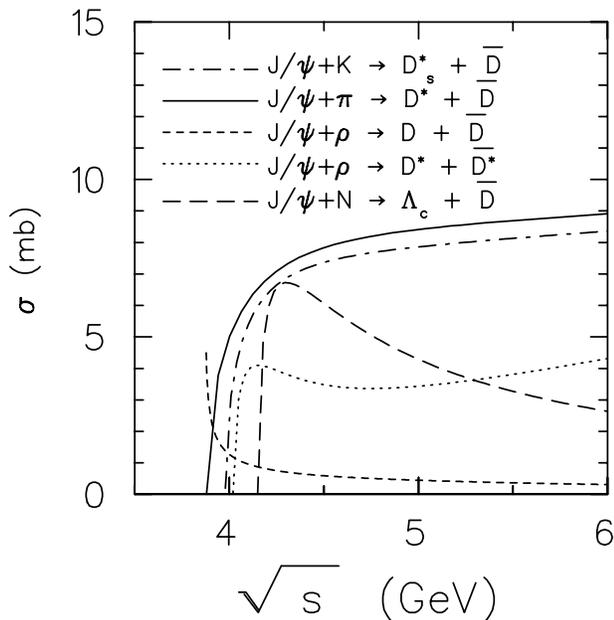}
    \caption{Dissociation cross sections for $\pi$, $\rho$, and kaon.  
             Nucleon-induced dissociation is also included.  See text
             for details.}
    \label{f2}
  \end{center}
 \end{figure}

Results for cross sections as functions of $\sqrt{s}$ are presented in 
Fig.~\ref{f2}.  
A startling feature in the endothermic reactions is 
the sharp rise just above threshold.   The cross sections reach maximum
strength a few hundred MeV above their respective thresholds.  
We note the significant difference between these results
and the $\pi$- and $\rho$-induced reactions from Ref.~\cite{Ma_98}, which
are due to the inclusion of contact terms and interference
effects.  Further, we note the relatively large kaon cross section.

Next we turn to the baryons and begin by writing an effective
lagrangian describing the interaction with pseudoscalars.  The
full {\it SU\/}(4) expression is quite lengthy\cite{DoGoHo}, so we 
concentrate on the particular term allowing analysis of the reaction
{\it J\/}/$\psi$ + {\it N\/} $\rightarrow \Lambda_{c}$ + 
$\bar{\it D\/}$.  The lagrangian bearing pseudoscalar coupling is\cite{Ko98}
\begin{eqnarray}
\mbox{\l}_{\,D\,N\,\Lambda_{c}} & = & g_{DN\Lambda_{c}}\left(
\bar{N}\gamma_{5}\Lambda_{c}\bar{D}+\bar{\Lambda}_{c}\gamma_{5}ND
\right).
\end{eqnarray}
The coupling constant is constrained by QCD sum rules which predicts
a value ${\it g}_{D\/N\/\Lambda_{c}}$ = 6.7 $\pm$ 2.1\cite{Nav_98}.  
There are two Feynman graphs, a direct and exchange contribution,
which when added together once again produces an amplitude 
respecting gauge invariance.  The cross section is presented also in
Fig.~\ref{f2}.  It to rises sharply at threshold to $\approx$ 7 mb
and falls with rising energy.

After the early stages of a high-energy heavy-ion 
reaction, the produced hot and dense system
rapidly expands and cools leaving a
kinetically thermal and probably even chemically 
equilibrated hadronic fireball.
The fireball quickly cools, most likely falls out of 
chemical equilibrium\cite{Koch,PrHa_99}, until 
it eventually freezes out.
Therefore, charmonium kinetics over a range of temperatures are needed
for establishing predictions for {\it J\/}/$\psi$ production.
Relativistic kinetic theory allows for simple
estimates of rates.  Nonequilibrium effects are of course
important, and will be discussed elsewhere\cite{kh99}.  In general,
the invariant four rate for {\it a\/} + {\it b\/} scattering assuming 
Boltzmann distributions is
\begin{eqnarray}
{\it dR\/} & = & \frac{g\ T\/^{\,2}}{(2\pi)^{\,4}}\int_{z_{\rm 
min}}^{\,\infty} {\it dz\/}\,\lambda(z^{\,2}T^{\,2},m_{a}^{2},m_{b}^{2})
{\it d\/}\sigma {\it K\/}_{1}(z),
\end{eqnarray}
where {\it g\/} is an overall degeneracy factor, {\it z\/} = $\sqrt{s}/T$,
{\it z\/}$_{\rm min}$ = max({\it m\/}$_{a}$+${\it m\/}_{b}$,{\it M\/})/{\it 
T\/}, with {\it M\/} being the sum of all final state particles' masses, and 
${\it K\/}_{1}$ is a modified Bessel function.    The number of times 
particle {\it a\/} scatters with a particle of type {\it b\/} per unit time 
is then
\begin{eqnarray}
{\rm Rate} & = & \frac{dR}{dN_{b}},
\end{eqnarray}
where ${dN_{b}}$ is the number density of {\it b\/} particles,
\begin{eqnarray}
{\it dN\/}_{b} & = & \frac{{\it g\/}_{b}}{\rm 2\pi^{\,2}}
{\it T\/}{\it m\/}^{\,2}_{b}
{\it K\/}_{2}({\it m\/}_{b}/{\it T\/}).
\end{eqnarray}
Again, ${\it K\/}_{2}$ is a modified Bessel function.

\begin{figure}[htbp]
  \begin{center}
  \epsfxsize 80mm 
  \epsfbox{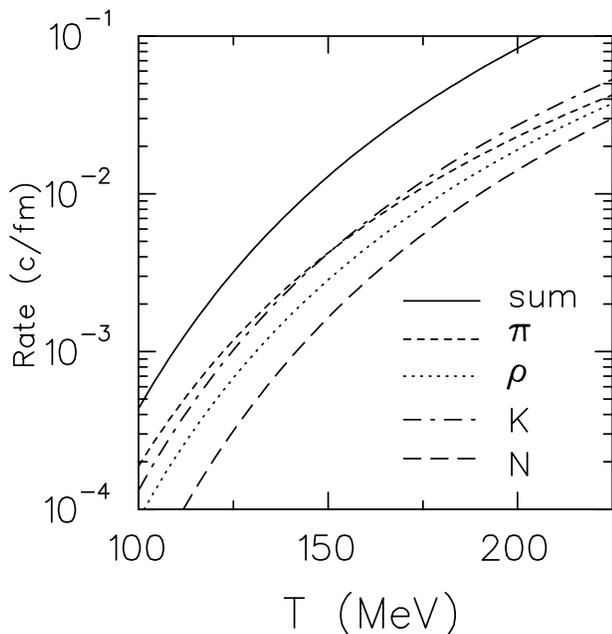}
    \caption{The thermal rates for {\it J\/}/$\psi$ dissociation
             induced by $\pi$ (short-dashed curve), $\rho$ mesons, (dotted
             curve), kaons (dot-dashed curve), and nucleons (long-dashed
             curve), and the sum (solid curve).}
    \label{f3}
  \end{center}
\end{figure}

The number of {\it J\/}/$\psi$ dissociations
per unit time induced by each light hadronic species are shown
separately in Fig.~\ref{f3} as well as a combined sum. 
{\it J\/}/$\psi$ dissociation rate in resonance matter is
$\approx$ 0.03 (fm/{\it c\/})$^{-1}$.
If we look toward temperatures approaching 200 MeV, the rate approaches 0.1 
(fm/{\it c\/})$^{-1}$.
It was previously thought that thermal hadronic dissociation rates were
so small as to be insignificant on the time scale of the fireball
created in heavy ion reactions.  Fig.~\ref{f3} indicates that 
reactions with light hadrons will indeed be an important hindrance 
for charmonium production.

There are of course error bars to discuss associated with these calculations. 
First, the coupling constants are uncertain to a level of ten of percent or so.
Form factors, which would tend to reduce the cross sections,
have not been included.   
On the other hand, a long list of reactions involving other
mesons could be imagined, and some of which might be important.
For instance, axial vector charm mesons {\it D\/}$_{1}$(2420)\cite{pdg} 
have relatively large widths and could therefore 
play an important role
in dissociating {\it J\/}/$\psi$ through such reactions
as
\begin{eqnarray}
\label{pijpsi}
\pi + {\it J\/}/\psi &\rightarrow & {\it D\/}_{1} + \bar{\it D\/}^{*},
\quad \bar{\it D\/}_{1} + {\it D\/}^{*}.
\end{eqnarray}
Another candidate likely affecting $J/\psi$ dissociation in the medium 
is $\chi_{c}$. 

Finally, the rate estimates
made here would be increased dramatically if phase space were overpopulated,
interpretable as finite chemical potentials.
As the fireball expands isentropically,
pion and kaon chemical potentials of order 50--100 MeV develop
in model calculations\cite{PrHa_99}.  This would easily buy
back a significant factor in the rates.  All these effects are currently
under investigation and will be reported upon separately\cite{kh99}.

\centerline{\rule{1.0in}{0.01in}}

It is a pleasure to thank Scott Pratt and Wolfgang
Bauer for valuable discussions during a visit to the National 
Superconducting Cyclotron Laboratory of Michigan State University where
my investigation began.  This work was supported in part by the National 
Science Foundation under grant number PHY-9814247.

\vskip -1\baselineskip

\end{document}